\documentclass[prl,twocolumn,showpacs,preprintnumbers,amsmath,amssymb]{revtex4}

\usepackage[dvips]{graphics}
\usepackage{tikz}
\begin{document}

\title{Is the fermionic exchange phase also acquired locally?}

\author{Chiara Marletto and Vlatko Vedral}
\affiliation{Clarendon Laboratory, University of Oxford, Parks Road, Oxford OX1 3PU, United Kingdom and\\Centre for Quantum Technologies, National University of Singapore, 3 Science Drive 2, Singapore 117543 and\\
Department of Physics, National University of Singapore, 2 Science Drive 3, Singapore 117542}

\date{\today}

\begin{abstract}
\noindent We argue that the fermionic exchange phase could be detected by local means. We propose a simple experiment to test our idea. This leads us to speculate that there might be a deeper mechanism behind the notion of particle statistics in quantum physics that goes beyond the conventional argument based on the spin-statistics connections. 
\end{abstract}


\maketitle                           %display desi d

Fermions are sometimes said to behave non-locally because swapping two fermions results in an extra phase factor in their joint state equal to $(-1)$. In terms of quantum fields, the fermionic operators acting on different modes anti-commute. Therefore, if one wants to enforce a tensor product structure on fermions (just like one does in the bosonic case), in order to preserve anti-commutation, one needs to introduce for each mode seemingly non-local operators. Specifically, the creation or annihilation operator acting on the $n$-th mode carries with it a phase that is equal to $\pi$ times the total number of fermions existing in modes 1 to $n-1$. This is the well-known trick of Jordan-Wigner employed in order to apply second-quantisation to fermions in the early days of quantum physics \cite{JW}. Nothing of this sort exists for bosons, since bosonic operators always commute for different modes. 

It is clear that the fermionic phase is not non-local in the sense that it does not allow one to signal faster than the speed of light. This kind of locality, sometimes called ``micro-causality", is built into the foundations of quantum field theory and any field, bosonic or fermionic, obeys it \cite{Weinberg}. 

However, we also know that phases in quantum physics are invariably acquired by local means. Namely, a particle that exists in a superposition of two different places $x$ responds to the local effective refractive index $n(x)$ in each place, leading to a phase factor of the form $\exp\big\{\frac{2\pi i n(x)}{\lambda} x\big\}$, where $\lambda$ is particle's de Broglie wavelength. The relative phase difference between two different paths $p_1$ and $p_2$ then becomes:
\begin{equation}
\exp\Bigg\{\frac{2\pi i (\int_{p_1} n(x)dx-\int_{p_2} n(x)dx)}{\lambda} \Bigg\}
\end{equation}
and is clearly observable by the means of interference. We briefly mention two different instances of this, though, as we said, all quantum phases are of this character. In a Mach-Zehnder interferometer light takes two different paths whose phase could be affected by wave plates that are inserted into these paths. For instance, a half wave plate in one arm of the interferometer changes the phase by $\pi$ with respect to the other arm. In a neutron interferometer, the same is true for individual neutrons, namely that different paths that the neutron takes could acquire different phases. One instance of the local refractive modulation in the neutron case could be gravity, since two paths separated by height $h$ would develop a phase difference of $\frac{m_N g h t}{\hbar}$, where $m_N$ is the neutron mass, $g$ is Earth's gravity and $t$ is the time of interference. This was tested in the pioneering COW experiment \cite{COW} and many times thereafter. 

Now, we would like to ask whether the fermionic exchange phase is of the same kind -- namely, also acquired by local means.  First of all, one might think that the fermionic phase is a global phase and therefore unobservable. But this would be a mistake.  We can use a simple quantum gate called a controlled swap (whose classical version is known as the Fredkin gate), to demonstrate that the phase is observable. Namely, we need an extra degree of freedom equivalent to a qubit, which will coherently control whether the two fermions are swapped, as in Figure 1. If the qubit is in the state $|0\rangle$, then no swap of fermions occurs, while if the qubit is in the state $|1\rangle$, the two fermions are swapped. The extra phase between the two then becomes the phase between the qubit states $|0\rangle$ and $|1\rangle$ and could therefore be detected. An implementation of this kind was recently proposed with cold atoms \cite{Roos} where the position degree of freedom played the role of the control qubit (this is particularly convenient as it does not require any additional systems to the two particles, although of course the extra qubit could in principle be any other physical system). Quantum control swaps have a wide range of uses in quantum information \cite{Andy} and computation \cite{Jayne,vanEnk}.
\begin{figure}[htb]
\centering
\includegraphics[scale=0.3]{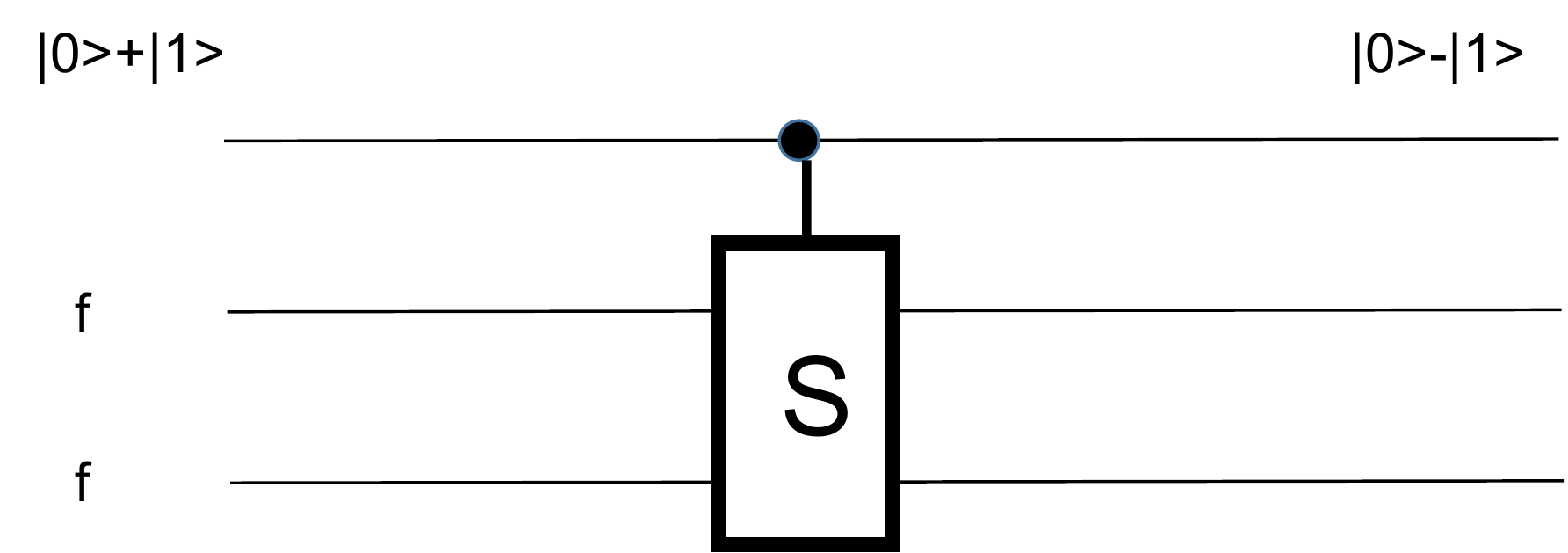}
\caption{A fermionic controlled swap gate. The control qubit is in a quantum superposition; the controlled systems are two fermionic particles, say electrons.}
\end{figure}

We note in passing that the phase would exist even if the fermions were completely distinguishable such as belonging to different species (e.g. an electron and a positron) providing that we insist that their corresponding operators should still anti-commute. Also, the swap is different from the experiment where we scatter two fermions off each other and observe the subsequent anti-bunching behaviour. If the two fermions were by any means distinguishable (say by their spin states or otherwise), then the scattering would always produce both the bunching as well as the anti-bunching behaviour. 

In order to address the question of locality we would like to perform the swap in a more continuous fashion. For that matter, suppose that we have a potential with $4$ sites as in the Figure 2a. Two fermions are initially in sites (spatial modes) $1$ and $3$. A swap is executed by first moving the fermions to sites $2$ and $4$ (say clockwise) and then moving them again in a clockwise fashion to come back to sites $1$ and $3$. A full swap is thus achieved in two steps and the minus phase is seen in the final state, as follows:
\begin{equation}
\textrm{step one}\;\;f^{\dagger}_4 f_3 f^{\dagger}_2 f_1  |1010\rangle = |0101\rangle
\end{equation}
\begin{equation}
\textrm{step two}\;\; f_4 f^{\dagger}_3 f_2 f^{\dagger}_1  |0101\rangle = -|1010\rangle
\end{equation}
where the first, second, third and fourth slot in the ket designate the first, second, third and fourth fermionic mode respectively. Note that, because of the anti-commutation relations, $f_2|0100\rangle = |0000\rangle$, but that $f_2|1100\rangle = -|1000\rangle$. This is precisely the non-local feature we refered to in the introduction, namely that the action on mode $2$, depends on the state of mode $1$. 
We should note in passing that we are not interested in the exact details of how this swap is performed. The above algebra is intended only to keep track of the fermionic phase, and we assume that any additional dynamical phase that is contingent of the details of the swap operation is under control and can be eliminated. 

Now, imagine that we do not swap the fermions, but instead interfere another two possibilities in which the first step has been done in the clockwise and counterclockwise directions simultaneously (as in Figure 2a). Even though this does not achieve a swap of fermions (each fermion has only gone half way), the two states will still acquire the same $\pi$ phase between each other. This is seen from the following simple algebra following operation:
\begin{equation}
\textrm{counterclockwise}\;\; f^{\dagger}_4 f_3 f^{\dagger}_2 f_1  |1010\rangle = |0101\rangle
\end{equation}
\begin{equation}
\textrm{clockwise}\;\; f^{\dagger}_2 f_3 f^{\dagger}_4 f_1 |1010 \rangle = - |0101\rangle
\end{equation}
This too will produce the same interference in the control qubit as when we perform the full controlled swap. Given that each electron traveled only half of the distance required for the full swap, it seems legitimate to argue that the phase was acquired by local means. 
\begin{figure}
\centering
\includegraphics[scale=0.3]{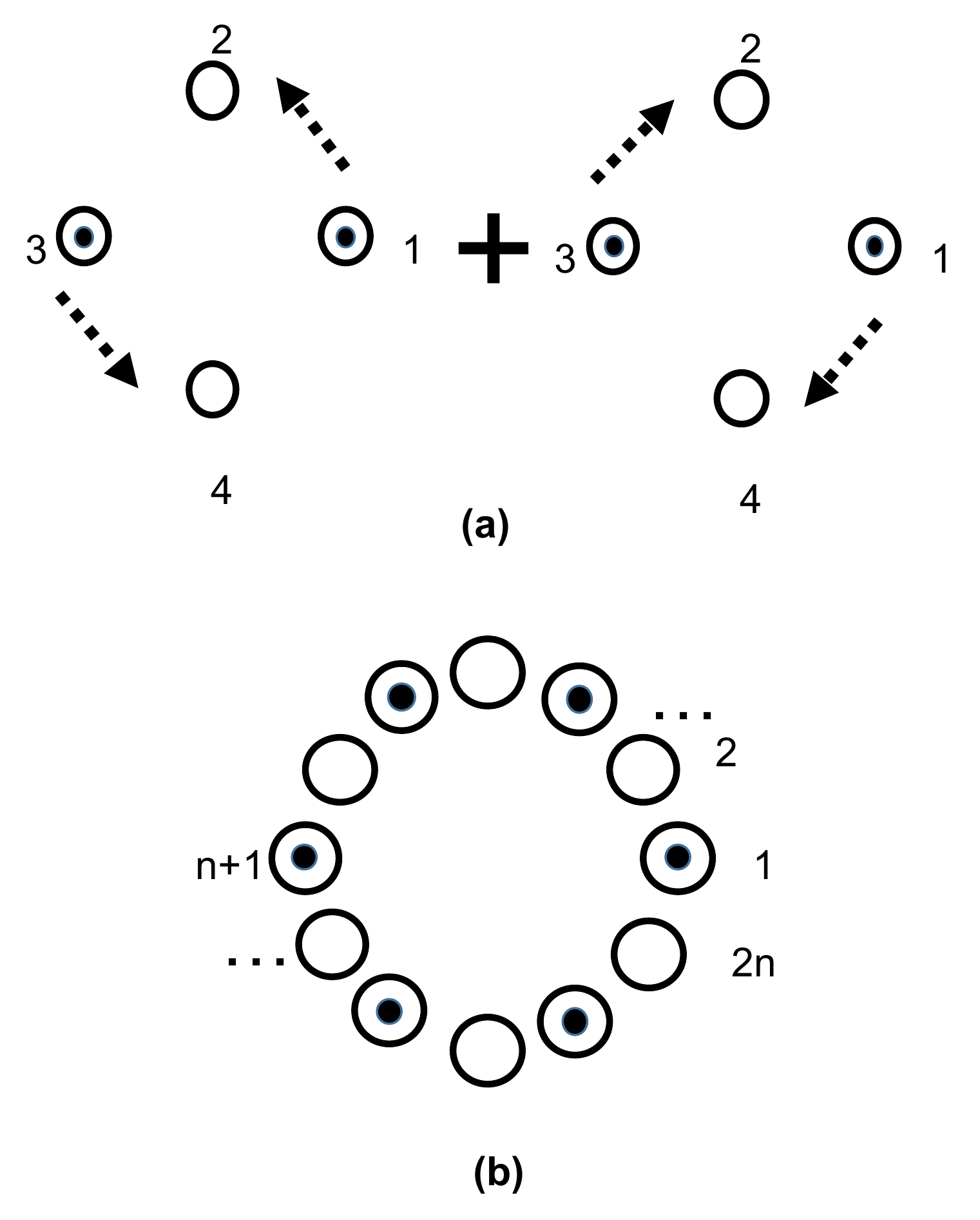}
\caption{(a): Pictorial representation of a controlled clockwise and counterclockwise partial swap of two fermions using four spatial modes. (b) Extension to n fermions placed in 2n spatial modes.}
\label{Figure1}
\end{figure}

In fact, this argument could be further refined. We could have $n$ fermions distributed evenly between $2n$ modes arranged in a circle (say that the fermions occupy only the odd modes as in Figure 2b).
Interfering the situations where they have made just one turn clockwise and one turn counterclockwise also produces the same fermionic phase. But now, none of the fermions ever moves more than just between the neighbouring sites (whose separation could be made arbitrarily small). This would seem to make the argument even stronger that the phase must be acquired by local means  - namely continuously along the swapping path.  If so, it highlights an interesting issue. Would this logic imply that there is a mediating field (of unknown origin) that is responsible for generating the exchange phase by coupling locally to fermions, in order to modify the local refractive index just like in the case of all other interference phenomena? 

Perhaps our argument points to the non-fundamental nature of fermions, which might ultimately be seen as composite bosonic particles or other emergent entities \cite{Wen}. We find this an intriguing possibility, especially that the fermionic phase is always the same, independently of the path taken when the swapping is implemented, and so whatever causes it would have to be of an intrinsically topological character.

\textit{Acknowledgments}: CM thanks the Templeton World Charity Foundation and the Eutopia Foundation. VV's research is supported by the National Research Foundation and the Ministry of Education in Singapore and administered by Centre for Quantum Technologies, National University of Singapore.


\begin{thebibliography}{1}
\bibitem{JW} P. Jordan and E. Wigner: Ober das Paulische Aquivalenzverbot. Z. Phys. 47, 631-651 (1928).
\bibitem{Weinberg} S. Weinberg, The Quantum Theory of Fields, (Cambridge University Press 1995). 
\bibitem{COW} R. Colella, A. W. Overhauser, and S. A. Werner, Observation of Gravitationally Induced Quantum Interference, Phys. Rev. Lett. {\bf 34}, 1472 (1975).
\bibitem{Roos} C. F. Roos, A. Alberti, D. Meschede, P. Hauke, and H. H\"affner, Revealing Quantum Statistics with a Pair of Distant Atoms, Phys. Rev. Lett. {\bf 119}, 160401 (2017).
\bibitem{Jayne} J. Thompson, M. Gu, K. Modi, and V. Vedral, Quantum Computing with Black-Box Subroutines, New J. Phys. 20, 013004 (2018).
\bibitem{Andy} Oscar C. O. Dahlsten, Andrew J. P. Garner, Jayne Thompson, Mile Gu, Vlatko Vedral,
Particle exchange in post-quantum theories, arXiv:1307.2529. 
\bibitem{vanEnk} S. J. van Enk, Exchanging identical particles and topological quantum computing, arXiv:1810.05208. 
\bibitem{Wen} M. Levin, and Xiao-Gang Wen, Photons and electrons as emergent phenomena, Rev. Mod. Phys. {\bf 77}, 871-879  (2005).




\end{thebibliography}
\end{document}